\documentclass[aip,rsi,reprint,graphicx]{revtex4-1}
\usepackage{graphicx}
\usepackage{subfigure}
\usepackage{hyperref}
\usepackage{amsmath}
\usepackage{amssymb}
\usepackage{epstopdf}
\usepackage{bbm}

\begin{document}

\title{Microcontroller-based locking in optics experiments}

\author{K. Huang}
\affiliation{
Laboratoire Kastler Brossel, UPMC Univ Paris 06, Ecole Normale Sup\'{e}rieure, CNRS, Coll\`ege de France, 4 place
Jussieu, 75252 Paris Cedex 05, France}
\affiliation{State Key Laboratory of Precision Spectroscopy,
East China Normal University, Shanghai 200062, China}

\author{H. Le Jeannic}
\affiliation{
Laboratoire Kastler Brossel, UPMC Univ Paris 06, Ecole Normale Sup\'{e}rieure, CNRS, Coll\`ege de France, 4 place
Jussieu, 75252 Paris Cedex 05, France}

\author{J. Ruaudel}
\affiliation{
Laboratoire Kastler Brossel, UPMC Univ Paris 06, Ecole Normale Sup\'{e}rieure, CNRS, Coll\`ege de France, 4 place
Jussieu, 75252 Paris Cedex 05, France}

\author{O. Morin}
\affiliation{
Laboratoire Kastler Brossel, UPMC Univ Paris 06, Ecole Normale Sup\'{e}rieure, CNRS, Coll\`ege de France, 4 place
Jussieu, 75252 Paris Cedex 05, France}

\author{J. Laurat}
\email{julien.laurat@upmc.fr}
\affiliation{
Laboratoire Kastler Brossel, UPMC Univ Paris 06, Ecole Normale Sup\'{e}rieure, CNRS, Coll\`ege de France, 4 place
Jussieu, 75252 Paris Cedex 05, France}


\begin{abstract}
Optics experiments critically require the stable and accurate locking of relative phases between light beams or the stabilization of Fabry-Perot cavity lengths. Here, we present a simple and inexpensive technique based on a stand-alone microcontroller unit to perform such tasks. Easily programmed in C language, this reconfigurable digital locking system also enables automatic relocking and sequential functioning. Different algorithms are detailed and applied to fringe locking and to low- and high-finesse optical cavity stabilization, without the need of external modulations or error signals. This technique can readily replace a number of analog locking systems advantageously in a variety of optical experiments.
\end{abstract}

\maketitle

\section{Introduction}

Optical setup usually relies on various locking systems for actively stabilizing interferometer paths or cavity lengths. Traditionally, a technique, such as, among others, the Pound-Drever-Hall \cite{PDH} (PDH), the dither-and-lock \cite{dither} or the tilt-locking \cite{tilt} processes, provides an error signal, which is then fed into an analog proportional-integral-derivative (PID) controller. In our quantum optics lab, for example, cavities containing non-linear crystals, i.e. optical parametric oscillators \cite{Morin2012,morinmode,jove}, or empty cavities for spatial-mode cleaning and noise filtering are commonly used \cite{Bachor,Keller}. Also, fringe locking is very often required when various beams are mixed or when light is injected into a parametric amplifier as the gain depends on the relative phase between the pump and this seed \cite{Morin2014}. Beyond the stability of the lock, other features can be very useful in these implementations: automatic searching of the locking point (especially for cavity peak locking), automatic relocking when the lock is lost, sequential locking by a sample-and-hold behavior. These practical features, which can enable for instance long-term data acquisition or just simpler functioning on a daily-basis, can be more easily implemented using digital systems.

Digital locking in optics experiments has indeed attracted more and more attention since the end of the nineties. It has been used for the first time in laser frequency stabilization \cite{AholaRSI1998,ZhaoRSI1998} and various implementations have been subsequently developed \cite{Rossi02,Pantelic03,Matsubara05,Seymour10}. Very recently, the fast development of high-speed and low-cost field programmable gate arrays (FPGA) and microcontroller units (MCU) opened a wealth of possibilities. For instance, in Ref. \onlinecite{SparkesRSI2011}, a FPGA programmed with National instruments LabView software is used to generate the modulation signal applied to an electro-optic modulator in a PDH scenario. The FPGA enables automatic relocking and includes some inbuilt locking analysis tools. Similarly, in Ref. \onlinecite{DietrichArXiv2009}, a microcontroller unit is also used in such a scenario to lock a fiber laser to an optical cavity.  In these implementations, the logic unit is usually combined with traditional locking processes. It enables the digitalization and processing of the signal, replacing the usual PID control and lock-in-amplifier, and adds novel capabilities and further scalability.

Here, we present a microcontroller-based locking system that does not require any modulation or external error signal. The core process is based on a maximum (or minimum) -searching algorithm, making the way to lock simple and intuitive. Additionally, the cost is extremely reduced as no other components, either optical or electrical, are required. We will detail the implemented algorithms and show the performances obtained for the locking of a Mach-Zehnder interferometer and of optical cavities with a moderate finesse (100) and a high finesse (1000). The performances are comparable to what can be obtained with typical analog lockings. 

Among the  large variety of microcontroller units now available, we chose the ADuC7020 from Analog Devices. Advantageously, this chip already contains five 12-bit ADCs (with sampling rate of 1MHz) and four 12-bit DACs (with voltage output settling time of 10 $\mu$s) \cite{ADuC7020}. Furthermore, the commercially-available evaluation board (EVAL-ADUC7020QSZ, around 50 euros) simplifies its use as no additional wiring is required.

There are ongoing discussions between choosing a FPGA or a conventional MCU for locking purpose. Generally speaking, FPGA can be much faster. However, for many lockings, fast digital controller is not necessary since the frequency bandwidth of the system is usually limited by the sensors, actuators and/or transducers.  For instance, the bandwidth of typical piezoelectric transducers (PZT) used for controlling the optical path length is only tens of kHz, which makes FPGAs largely exceeding the needs. Last but not least, microcontrollers can be programmed in C language, which makes them easily accessible by relatively inexperienced users.

\section{Locking a Mach-Zehnder interferometer}\label{max}
As a first implementation, we consider the case of a Mach-Zehnder interferometer with 1.5 meter-long arms, as depicted in Fig. \ref{figure1}(a). One photodiode is used to measure the output power and the signal is fed into the ADC input of the microcontroller unit. The MCU output is then sent to a piezoelectric transducer (PI P-016.00H) to control the optical length of one arm. The algorithm used here is a minimum-searching approach with an additional automatic PZT recentering, as detailed now. 

\subsection{Searching algorithm}

\begin{figure}[t!]
\centerline{\includegraphics[width=0.92\columnwidth]{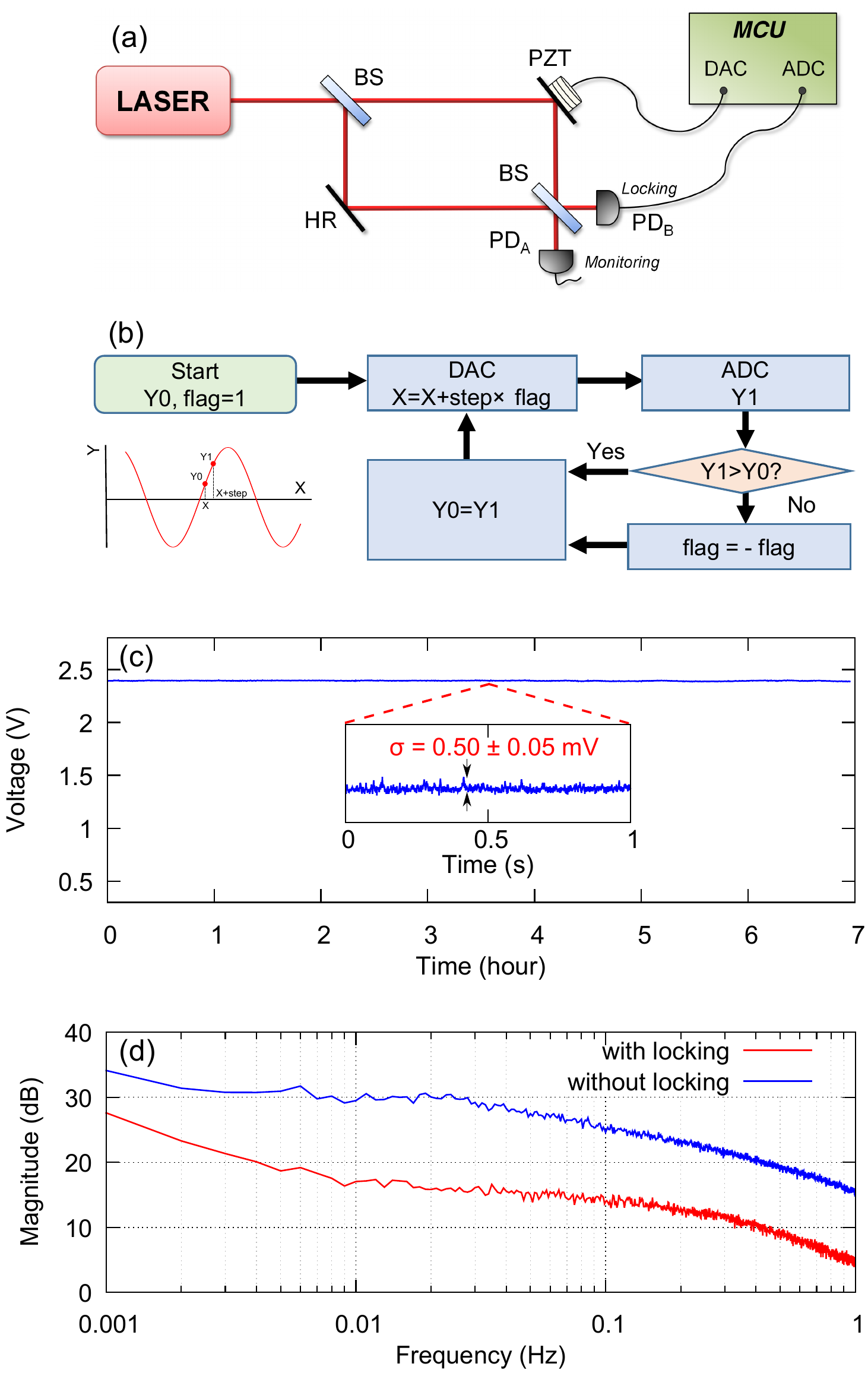}}
\caption{(Color online) (a) Experimental setup: locking a 1.5-meter-arm Mach-Zehnder interferometer with a microcontroller unit. HR stands for high-reflective mirror, BS for beam-splitter, PZT for piezoelectric transducer and PD for photodiode. (b) Flowchart of the program execution for the maximum-searching algorithm. (c) Long-term stability. The inset gives a one-second zoom and the associated standard deviation. (d) Noise spectrum at low frequencies with and without locking.}
\label{figure1}
\end{figure}

A flowchart of the program execution is given in Fig. \ref{figure1}(b). The voltage on the PZT is increased with a constant step unless the signal value (Y1) is smaller than the previous one, which would lead to a sign-flip of the sweep. As a result, the locked signal can always stay at its maximal value. To avoid disturbances induced by high frequency noise of the detector ($PD_B$) the signal is averaged over a user-defined numbers of ADC sampling (typically 50). 

Importantly, in such long-arm interferometers, fluctuations can lead to a change of the paths by many wavelengths, larger than the limited range of typical actuators. In order to keep the locking, one has thus to shift back the position of the PZT. Usual analog PID controllers do not have this kind of feature, and would need elaborated developments. In contrast, this can be easily achieved by a programmable device. The recentering can indeed be trivially taken into account with simple codes in the locking program. Specifically, when the detected peak position is near the ends of the DAC output range, the output is shifted back to the center (not shown on the flowchart). As a result, the phase can be automatically relocked around the center of the DAC range. After a certain time (usually a few milliseconds), a stable locking is recovered. Such automatic locking enables a very long-term stability, as shown in Fig. \ref{figure1}(c) for a 7-hour recording. Typically, recentering occurred once per hour. 

The inset in Fig. \ref{figure1}(c) corresponds to a one-second zoom. The standard deviation (with a sampling rate of 250 kHz) is $0.50\pm 0.05$ mV, which can be translated into a phase fluctuation of $1.7^{\circ}\pm 0.1^{\circ}$ ($1.5^{\circ}$ if corrected from detector noise). Over 1 hour, the standard deviation raises to $1.7\pm 0.3$ mV, corresponding to $3.1^{\circ}\pm 0.2^{\circ}$. The MCU-based locking enables thus a very good phase stability on the short as well as on the long time scales, compatible with most applications, including for instance the use of high squeezing levels \cite{Takeno07,Eberle13,jove} in demanding quantum optics and quantum information experiments.

\subsection{Simple model and set parameters}
A simple model can be used to estimate the locking performances based on this algorithm. The phase stability is indeed limited by the ADC resolution $\epsilon$ of the microcontroller. In our case, with a number of bits $N=$12 and a maximum reference voltage $V_0=2.5$ V, the resolution is given by $\epsilon=V_0/2^{12}\simeq 0.61$mV. Only the voltage fluctuations larger than $\epsilon$ can be detected, leading to a minimum detectable phase change $\Delta\theta_0$. 

To minimize this value, the best strategy is to use the full dynamic range, and thus having a fringe with close to unity visibility, i.e. a detected voltage $V=V_0\sin^2(\varphi/2)$. In this case, the minimum detectable phase change is given by:
\begin{eqnarray}
\Delta\theta_0\simeq 2\sqrt{\frac{\epsilon}{V_0}}=2\sqrt{2^{-N}}\simeq1.8^\circ.
\end{eqnarray}
For step sizes in the algorithm leading to phase changes smaller than this value, the locking noise, i.e. the measured standard deviation, is expected to be constant, with a value slightly smaller than $\epsilon$. In the general case of a non-unit visibility, $V_0$ has to be replaced by the fringe amplitude (2.2 V in our case). One way to increase the locking performance would be to combine two 12-bit ADCs to obtain a 24-bit system \cite{DietrichArXiv2009}. Another possibility is to use a logarithmic amplifier in order to increase the resolution around the set point.

We now consider step sizes leading to phase changes larger than the previous one. The resulting phase increment depends on the number of bits $N$ of the DAC, the number $n$ of fringe periods for a full scan and an integer $M$ defined by the user:
\begin{eqnarray}
\Delta\theta  = M . \frac{2\pi n}{2^N}.
\end{eqnarray}
When the phase step $\Delta \theta$ is larger than $\Delta {\theta _0}$, the phase fluctuation scales with $\Delta \theta$. The noise is driven by the steps. For proper functioning, this regime should be avoided.  

 \begin{figure}[t!]
\centerline{\includegraphics[width=0.93\columnwidth]{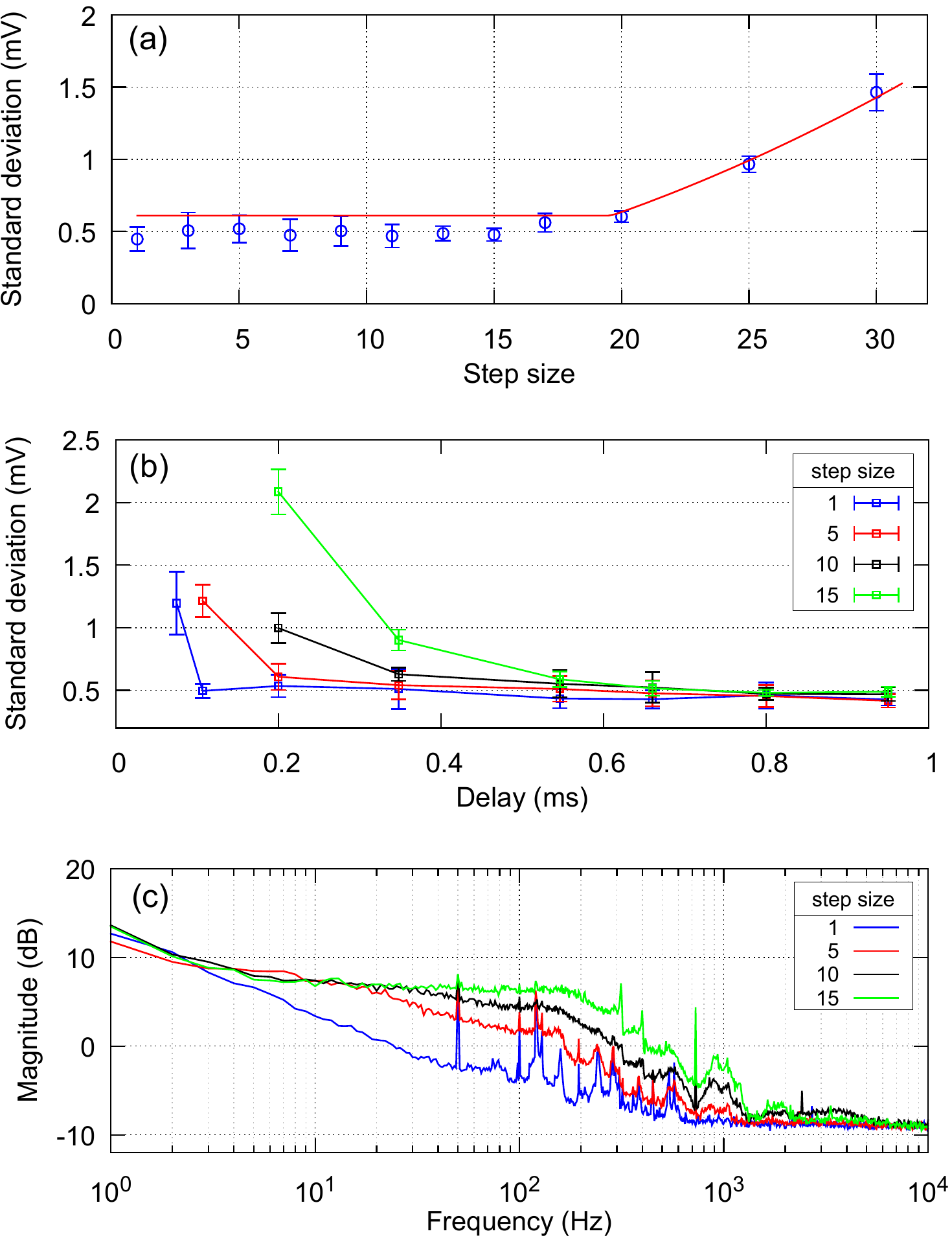}}
\caption{(Color online) (a) Standard deviation as a function of the step size given by the integer $M$. The solid line is a guide for the eyes, following the model given in the text. (b) Standard deviation as a function of the delay between the output (DAC) and the acquisition (ADC). (c) Noise spectrum for different step sizes. The delay is set to 800 $\mu$s. }
\label{figure2}
\end{figure}

For a full characterization of the locking system, and to illustrate this simple model, we  studied in a systematic way the behavior obtained for various parameters in the algorithm. Figure \ref{figure2}(a) shows the standard deviation for one second as a function of the step size. As predicted, the measured voltage noise exhibits first a plateau and then scales as the square of the step. If the step size is increased further, a stable locking cannot be obtained anymore.

One parameter more difficult to include in this simple model is the temporal delay between the output (DAC) and read (ADC) blocks. With a fixed step size, the delay plays an important role in the locking behavior as shown in Fig. \ref{figure2}(b). Specifically, when the delay is too long, the bandwidth is too much reduced and cannot compensate for the fluctuations. When the delay is too short, there is not enough time for the PZT to respond. Because the locking algorithm relies on the comparison of the two sequential signal values, the improper response of PZT will produce some disturbance for each step, finally altering the locking. A 1 ms-delay is typically added in the program. A bandwidth around a few kHz is achieved here, limited by the PZT. 

Finally, Figure \ref{figure2}(c) provides the noise spectra for different step sizes. The plateau above 3 kHz is due the limited vertical resolution (8 bits) of the oscilloscope used for the data acquisition.

\section{Locking an optical cavity}
The microcontroller-based locking is now applied to a more complicated task, i.e. the locking of optical cavities. Two cavities with moderate and high finesse, respectively around 100 and 1000, will be considered, requiring two different execution programs. 

\subsection{Low-finesse cavity}
\begin{figure}[t!]
\centerline{\includegraphics[width=0.93\columnwidth]{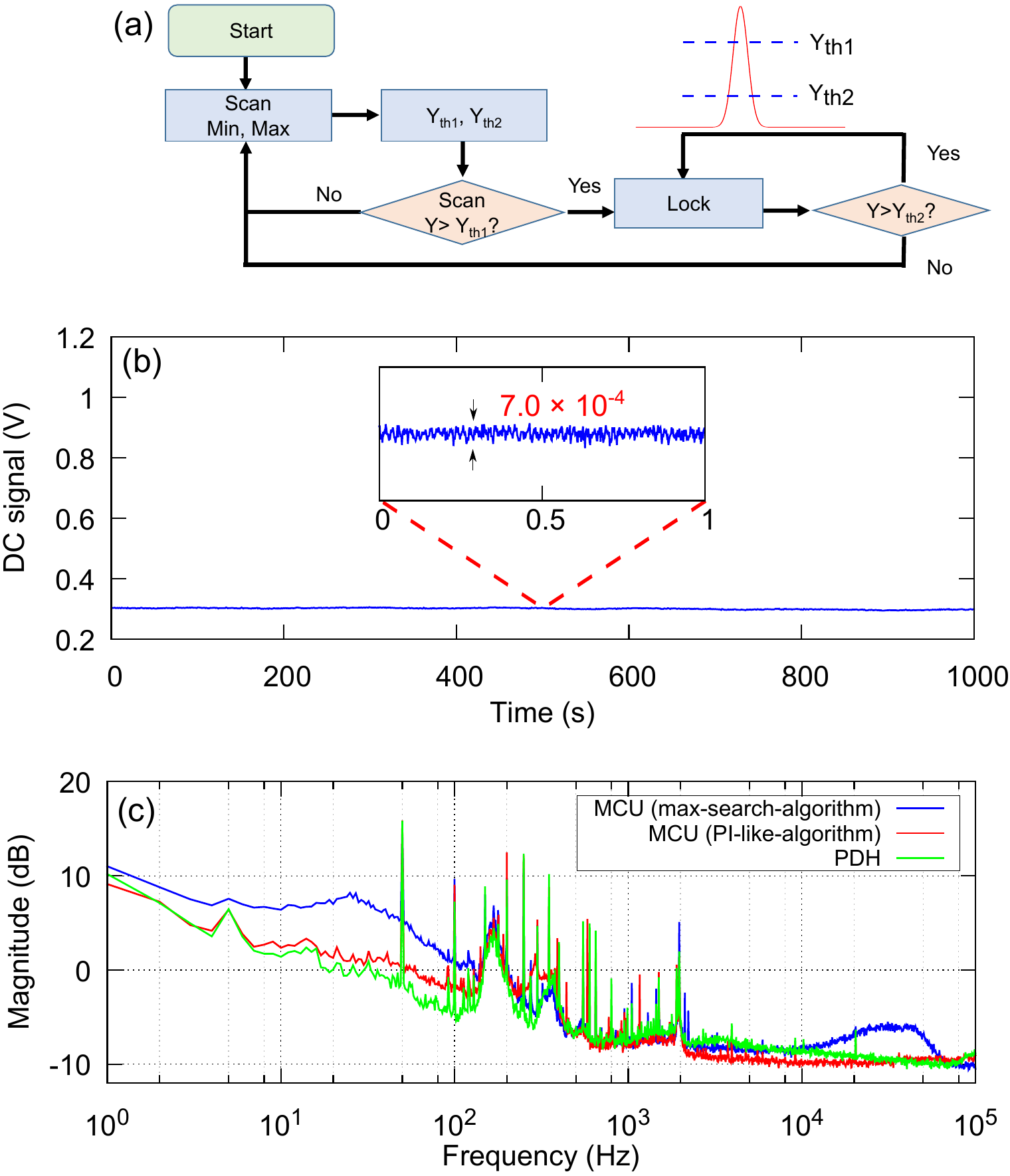}}
\caption{(Color online) (a) Flowchart of the program execution to lock a low-finesse cavity, including peak searching and automatic relocking. (b) Long-term stability. The inset gives a one-second zoom. The standard deviation is normalized to the cavity peak height. (c) Noise spectra for maximum-searching locking, PI-like locking and for locking with the standard analog Pound-Drever-Hall technique.}
\label{figure3}
\end{figure}

The locking of a cavity specifically requires a scanning mode to first identify the peak heights and positions of the cavity resonance. This function can be easily realized with simple programming for the generation of triangle signal. The flowchart of the program execution is given in Fig. \ref{figure3}(a). The microcontroller first sweeps the cavity length and subsequently defines a high and low threshold ($Y_{th1}$ and $Y_{th2}$). It then sweeps again the length to reach an initial start point above the high threshold. If the locked signal becomes smaller than the low threshold, the microcontroller will get out of the locking mode and go back to the scanning mode. 

To test this method, we used an optical parametric oscillator, i.e. a cavity with a non-linear crystal inserted into it \cite{jove}. The finesse for the pump is around 100 and the 4 cm-long cavity was previously locked by the PDH technique, with a phase-modulation at 12 MHz. Figure \ref{figure3}(b) confirms the long-term stability and the inset shows a one-second zoom. The standard deviation normalized to the peak height is equal to $7.0\times10^{-4}$, similar to what is obtained here with the PDH locking ($6.6\times10^{-4}$). For 15 minutes, the normalized standard deviation raises to $4\times10^{-3}$.

To obtain this result, the locking algorithm has been slightly modified. Indeed, with the maximum-searching algorithm presented previously in Sec. \ref{max}, the standard deviation is a bit larger ($8.0\times10^{-4}$). This simple approach is however sufficient for many applications. Here, the control has been improved by implementing some proportional feedback using more sophisticated programming. For instance, one can numerically calculate the corresponding derivative of the signal, or simply use the signal difference between the sequential steps, as the gain of the feedback signal. We call this method the PI-like algorithm. The measured noise spectra are finally given in Fig. \ref{figure3}(c) for the three different techniques. 

\begin{figure}[t!]
\centerline{\includegraphics[width=0.93\columnwidth]{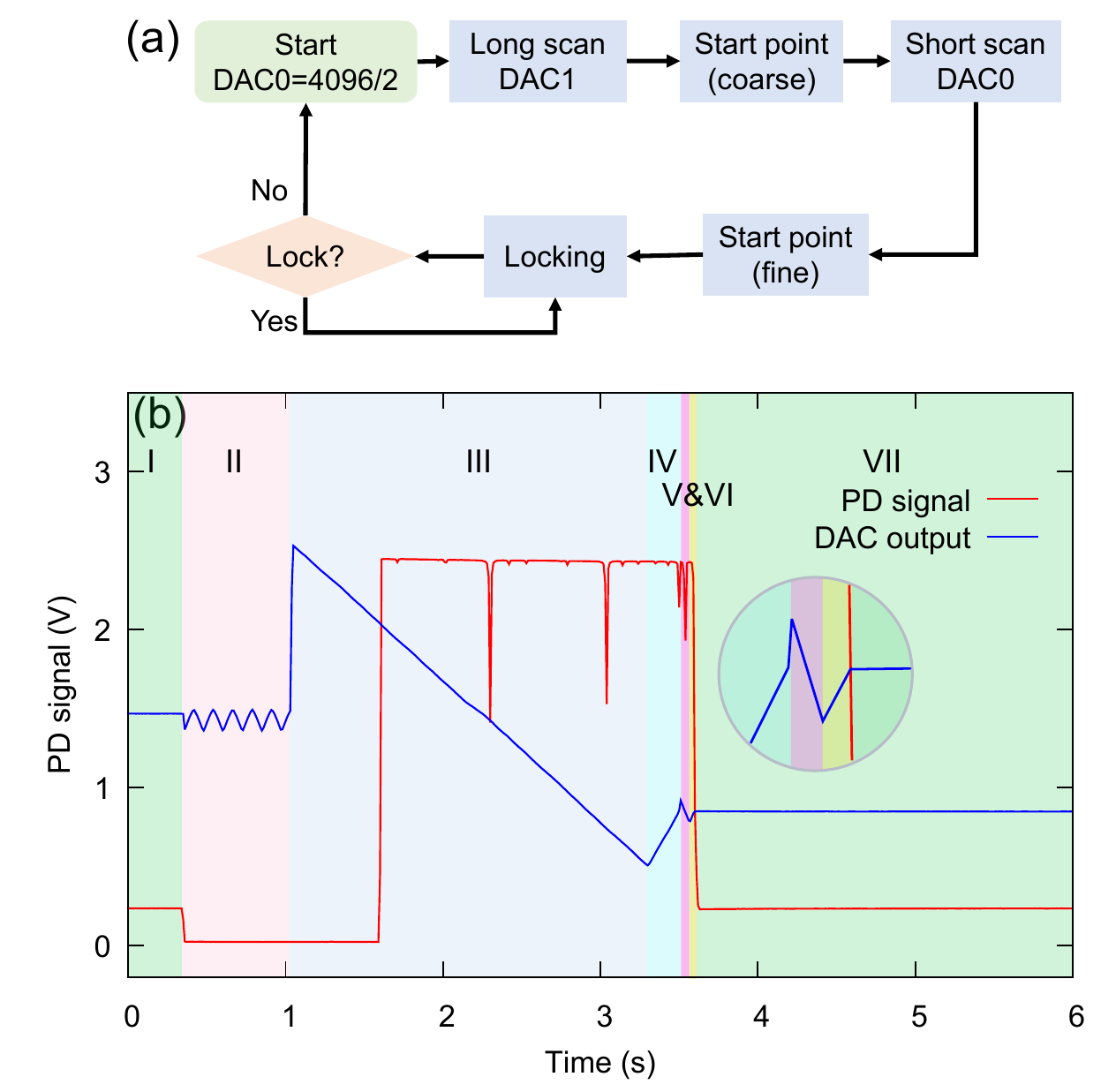}}
\caption{(Color online) (a) Flowchart of the program execution to lock a high-finesse cavity. This scheme includes two different scanning modes controlled by two outputs in order to use the full DAC resolution. (b) Experimental locking process of a cavity with a finesse equal to 1000. The light is measured in reflexion. (I) Cavity is locked. (II) Light is blocked to disrupt the lock and the program goes thus to the short-scan mode. (III) As the short-scan did not enable the relocking in this specific example, the long-scan is started in order to learn the rough peak position. (IV) The output is set at a rough start point determined in the previous step. (V) A short-scan is started to find the precise peak position. (VI) The output is set to the precise start point. (VI) Locking is on.}
\label{figure4}
\end{figure}

\subsection{High-finesse cavity}
Compared to a low-finesse cavity, locking a high-finesse cavity is more challenging due to the limited resolution of the DAC. For instance, for a finesse of 1000, the number of sampling points to cover the whole peak is around $4096/1000\simeq 4$ which is obviously insufficient to lock the cavity. To overcome this problem, one can combine two 12-bit DACs in order to obtain a 24-bit ones as mentioned previously. Another way to address the problem is to use two different scanning modes, namely a long scan and a short scan, as explained now.

The flowchart of the program execution is given in Fig. \ref{figure4}(a). One output of the DAC (DAC1) is used for the long scan, which spans over more than a free spectral range in order to identify the rough peak position. Another output (DAC0) is used for the short scan around the peak position identified by the long scan mode and also used for the locking. The two outputs are summed with different gains. Such uneven allocation of sampling over the free spectral range enables the full use of the DAC resolution. Since there are four DACs available in the microcontroller development board, this method can be easily implemented without the need of any additional electronic building.

Figure \ref{figure4} gives an example of the process for locking a 0.5 mm-long cavity\cite{Morin2012} with a finesse of 1000. Note that in order to make the re-lock process faster when the cavity is unlocked, the program goes firstly to the short-scan mode instead of directly using the long-scan one. If the short scan still cannot find the peak after 10 sweeps then the program jumps to the long-scan mode. 

\begin{figure}[t!]
\centerline{\includegraphics[width=0.93\columnwidth]{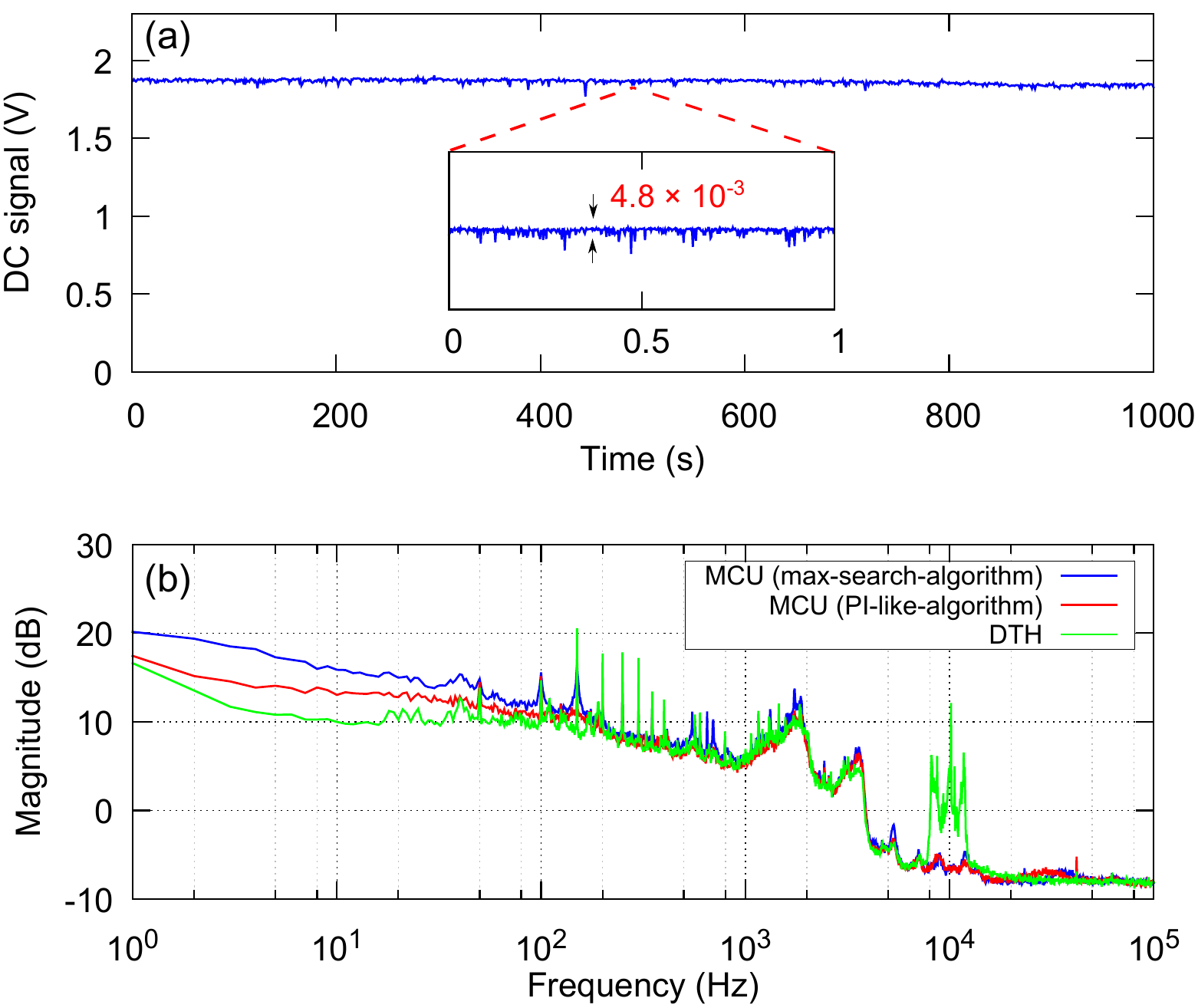}}
\caption{(Color online) (a) Long-term stability for the high-finesse cavity. The inset gives a one-second zoom. The standard deviation is normalized to the cavity peak height. (b) Noise spectrum for maximum-searching locking, PI-like locking and the traditional analog dither-and-lock technique.}
\label{figure5}
\end{figure}

The performances are displayed in Fig. \ref{figure5}. Figure \ref{figure5}(a) provides the temporal stability using the PI-like algorithm for the locking, while Fig. \ref{figure5}(b) gives the spectra for PI-like, maximum-searching and analog dither-and-lock (DTH) techniques. For one second, the standard deviation normalized to the peak height is found to be $4.8\times10^{-3}$ for the PI-like, which is slightly better than the $5.6\times10^{-3}$ given by DTH. The direct maximum-searching provides a standard deviation of $6\times10^{-3}$. For 15 minutes, the normalized standard deviation reaches $8\times10^{-3}$.
These results show the suitability and efficiency of the microcontroller-based locking with high-finesse cavities.

\begin{figure}[t!]
\centerline{\includegraphics[width=0.78\columnwidth]{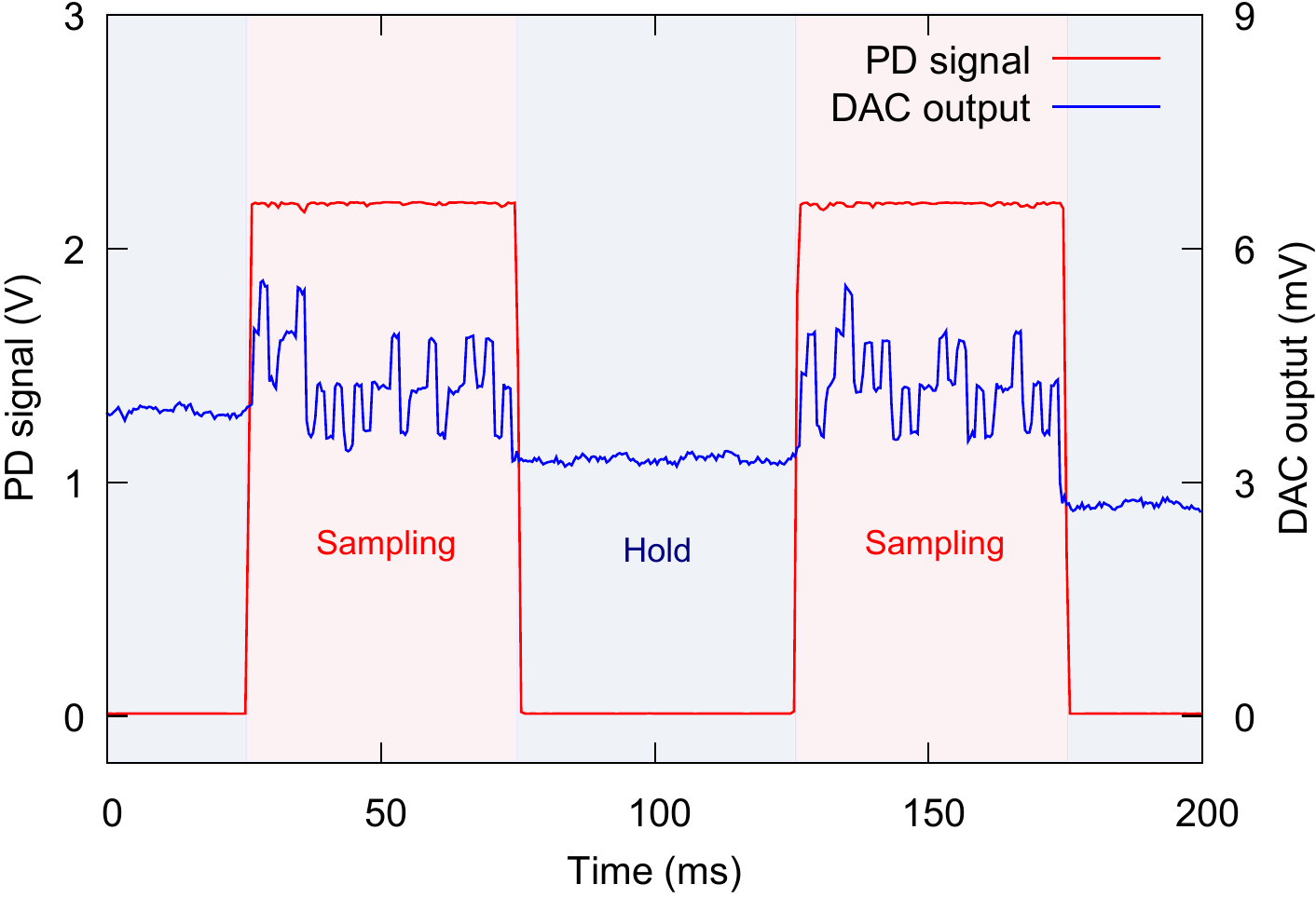}}
\caption{(Color online) Sequential locking for the high-finesse cavity. The light is switched off by mechanical shutters during 50 ms between locking periods. The DAC output holds the last value. This sequential functioning is very useful in many experiments where data acquisitions have to be done in the absence of strong lights, such as in experiments involving photon counting. }
\label{figure6}
\end{figure}

As an additional illustration of functions easily implemented by programming, figure \ref{figure6} finally shows the temporal trace obtained for a sequential locking. The cavity is locked for 50 ms and the light is then blocked. The unit keeps the last output value for a specified time, until the light is switched on again. This sample-and-hold functioning is central to many experiments: for instance, it enables to lock an experiment and then to proceed to acquisition in the photon-counting regime where locking lights have to be off \cite{Morin2014}.

\section{Conclusion}
In conclusion, we have shown that the currently available microcontroller units, such as the 12-bit ADuC7020, can be used for various optical lockings, including high-finesse cavities. It can advantageously replace traditional analog systems, offering similar performances in the presented examples but also additional features that are easy to tailor by C programming. In complex experiments where a great number of lockings are involved, this practical digital control also enables for instance delayed locking when systems are cascaded or very long data-acquisition. Besides their use in laser-based researches, these microcontroller systems can also be easily implemented in undergraduate educational physics laboratory as a good introduction to feedback and control systems.

\begin{acknowledgments}
This work is supported by the ERA-Net CHIST-ERA (QScale), the Emergence program from Ville de Paris (NanoQIP) and the European Research Council Starting grant HybridNet. K. Huang acknowledges the support from the Foundation for the Author of National Excellent Doctoral Dissertation of China (PY2012004) and the China Scholarship Council. J. Laurat is a member of the Institut Universitaire de France.
\end{acknowledgments}

\end{document}